\documentclass[%
reprint,
 amssymb, amsmath,%
 aps,
 prl,
]{revtex4-1}

  \usepackage[pdftex]{graphicx}
  \graphicspath{{../figures/}}
  \DeclareGraphicsExtensions{.pdf,.jpeg,.png}

\usepackage{subeqnarray}
\usepackage{stmaryrd}
\usepackage{docs}%
\usepackage{bm}%
\usepackage[colorlinks=true,linkcolor=blue]{hyperref}%
\expandafter\ifx\csname package@font\endcsname\relax\else
 \expandafter\expandafter
 \expandafter\usepackage
 \expandafter\expandafter
 \expandafter{\csname package@font\endcsname}%
\fi
\begin{document}

\title{%
       Characterization of Zero-Bias Microwave Diode Power Detectors \\at Cryogenic Temperature.}

\author{Vincent Giordano$^{1}$}%
\email{giordano@femto-st.fr}
\author{Christophe Fluhr$^{1}$, Beno\^{i}t Dubois$^{2}$ and Enrico Rubiola$^{1}$}

\affiliation{\vskip 5mm $^{1}$ FEMTO-ST Institute - UMR 6174\\
CNRS / ENSMM / UFC / UTBM\\
26 Chemin de l'\'Epitaphe\\
25000 Besan\c{c}on -- FRANCE}%

\affiliation{\vskip 5mm $^{2}$ FEMTO Engineering\\
32 avenue de l'Observatoire \\
25000 Besan\c{c}on -- FRANCE\\}%
\vskip 5mm
\date{2016 January, 25}%

\begin{abstract}
We present the characterization of commercial tunnel diode low-level microwave power detectors at room and cryogenic temperatures. The sensitivity as well as the output voltage noise of the tunnel diodes are measured as functions of the applied microwave power, the signal frequency being 10 GHz. We highlight strong variations of the diode characteristics when the applied microwave power is higher than few microwatt. For a diode operating at ${4}$ K, the differential gain increases from ${1,000}$ V/W to about ${4,500}$ V/W when the power passes from ${-30}$ dBm to ${-20}$ dBm. The diode present a white noise floor equivalent to a NEP of ${0.8}$ pW/ ${\sqrt{\mathrm{Hz}}}$ and ${8}$ pW/${ \sqrt{\mathrm{Hz}}}$ at 4 K and 300 K respectively. Its flicker noise is equivalent to a relative amplitude noise power spectral density ${S_{\alpha}(1~\mathrm{Hz})=-120}$~dB/Hz at ${4}$ K. Flicker noise is 10 dB higher at room temperature.
\end{abstract}

\maketitle

\section{Introduction}

Tunnel diodes are known to be very efficient low-level microwave power detectors. They are available off-the-shelf from numerous manufacturers in the form of small packaged coaxial components that can be easily implemented in any microwave system. They are used in power, AM noise measurement, control  and monitoring systems. Tunnel detectors are actually backward detectors. The backward diode is a tunnel diode in which the negative resistance in the forward-bias region is made negligible by appropriate doping, and used in the reverse-bias region.
Most of the work on such detectors dates back to the sixties \cite{burrus63,cowley66,gabriel67}. Tunnel detectors exhibit fast switching and higher gain than the Schottky counterpart. Tunnel diodes also work in cryogenic environment, provided the package tolerates the mechanical stress of the thermal contraction.\\

In this paper we present a characterisation of commercial tunnel diodes (Model Herotek DT8012). Such diodes are implemented in our 10 GHz ultra-stable Cryogenic Sapphire Oscillators (CSO)\cite{rsi10-elisa,RSI-2012}. They are used as low-level detectors to thinly control the phase and the power of the signal in the oscillating loop. In this way they are key components as their sensitivity conditions the CSO frequency stability. In this application the measurement bandwidth is limited to 100 kHz, we thus do not explore the switching capability of the tunnel diode and restrict our characterisation to the low frequency sensitivity.\\

Firstly, we introduce the topic of AM noise measurement. The tested diodes specifications and the corresponding electrical model are given in section \ref{sec:diode-spec}. Then we describe the experimental set-up and the power-to-voltage gain measurements (section \ref{sec:diode-carac}). The principle and the results of the diode intrinsic noise measurements are presented in section \ref{sec:diode-noise}. The application to AM noise control is given in section \ref{sec:appli}.\\

\section{AM noise measurement basics}
\label{sec:AM-measurement}

Our diode detectors are intented to measure the small power fluctuations affecting the signal of an ultra-stable oscillator. Considering only AM noise the RF voltage is:

\begin{equation}
V_{RF}(t)=V_{0}\left [ 1+\alpha(t) \right ] \cos(2\pi \nu_{0} t)
\label{equ:VRF}
\end{equation}

where $V_{0}$ is the RF voltage mean amplitude, $\nu_{0}$ is the signal frequency and $\alpha(t)$ is the fractional amplitude fluctuation. In low noise conditions, we have $\left | \alpha(t) \right | \ll 1$ and the instantaneous signal power is:

\begin{equation}
P(t)=P_{0}+\delta P(t) \approx \dfrac{V^{2}_{0}}{2R_{0}} \left ( 1+2\alpha(t) \right )
\label{equ:P}
\end{equation} 

where $R_{0}=50 \Omega$, $P_{0}$ is the mean signal power. The amplitude fluctuations are measured through the measurement of the power
￼fluctuation $\delta P(t)$,

\begin{equation}
\alpha(t)= \dfrac{1}{2} \dfrac{\delta P(t)}{P_{0}}
\label{equ:alpha}
\end{equation} 

and of its Power Spectrum Density (PSD),

\begin{equation}
S_{\alpha}(f)= \dfrac{1}{4} S_{\frac{\delta P}{P_{0}}}(f) = \dfrac{1}{4P^{2}_{0}} S_{\delta P}(f)
\label{equ:Salpha}
\end{equation}

\begin{figure}[ht!]
\centering
\includegraphics[width=\columnwidth]{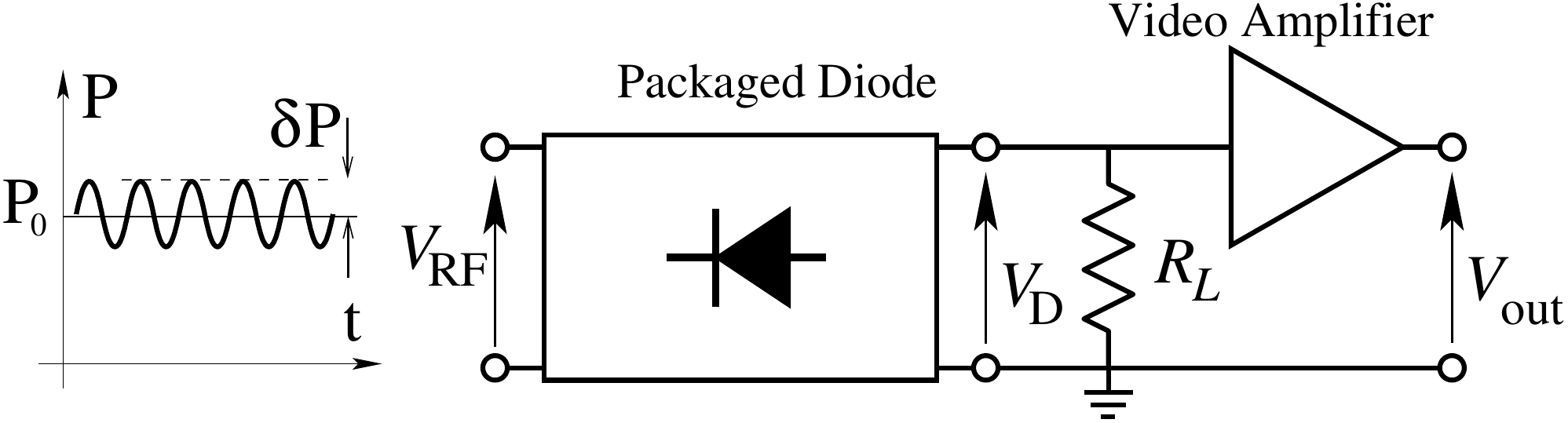}
\caption{The basic implementation of a zero-bias microwave power detector.}
\label{fig:fig1}
\end{figure}

The basic implementation of a zero-bias microwave power detector is represented in Fig. \ref{fig:fig1}. %

The RF voltage is applied to the diode input terminal. The video output is loaded by $R_{L}$ and the output voltage $V_{\!D}$ is a function of the input signal power $P$. A perfect square-law detector will produce a voltage  proportional to $P$:
\begin{equation}
V_{\!D}= K_{\!D} P
\label{equ:VD}
\end{equation}

$K_{\!D}$ is  the detector power-to-voltage gain expressed in V/W, also called \it{sensitivity}\rm\ in most the data-sheets. Manufacturers generally specify $K_{\!D}$ measured at room temperature and for an input signal power of $-20$ dBm or so. Typicaly for a X-band diode, $K_{\!D}\approx 1,000$ V/W. \\

It is generally admitted that equation \ref{equ:VD} holds for a signal power below $-20$ dBm and then the detector response turns smoothly from quadratic to linear. At high signal power, typically above $0$ dBm, the diode is acting as a peak detector.
This traditionally-accepted view of the diode detector leads to expect a sensitivity $K_{\!D}$ almost constant below $-20$ dBm and then saturating smoothly when the signal power is increased like in \cite{locke08}. As we will see in the following this simple model is not sufficient to describe the diode behavior on its whole dynamic range. \\

Providing equation \ref{equ:VD} holds and knowing $K_{\!D}$, $S_{\alpha}(f)$ is easily determined through the measurement of the detector output voltage PSD. However the intrinsic detector noise will limit the measurement of the signal power fluctuations.
Looking at the specifications of commercial power detectors, information about noise is scarce. Some manufacturers give the NEP (Noise Equivalent Power) parameter, i.e., the power at the detector input that produces a video output equal to that of the device noise. In no case is said whether the NEP increases or not in the presence of a strong input signal, which is related to precision. Even worse, no data about flickering is found in the literature or in the data sheets.

\section{Diode specifications and electrical model}
\label{sec:diode-spec}
\begin{figure}[h]
	\centering
	\includegraphics[width=\columnwidth]{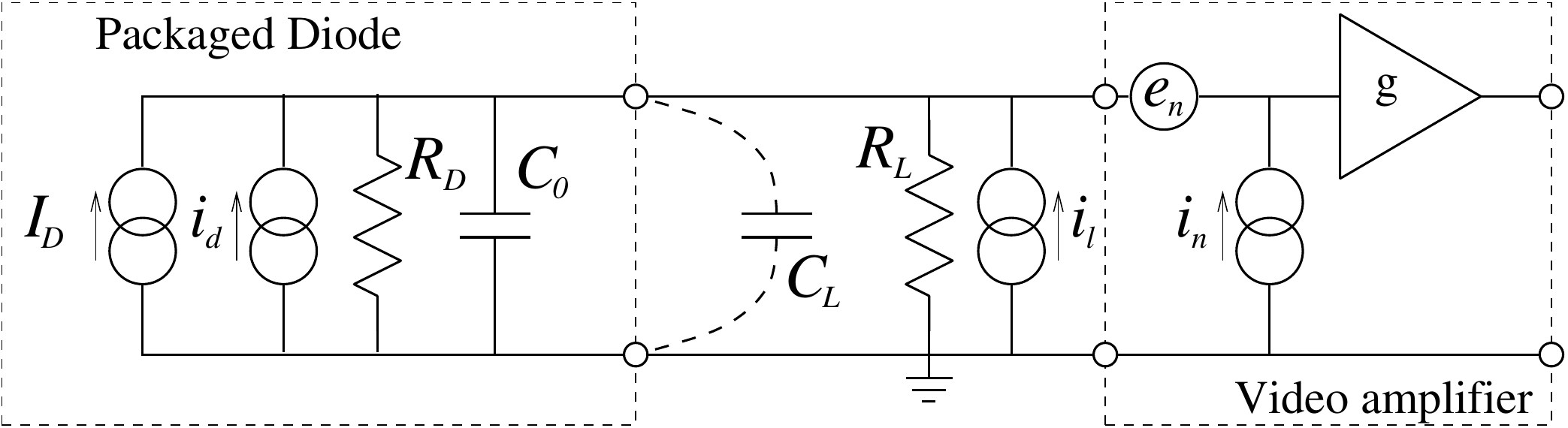}
	\caption{Electrical model for the detector video output. The capacitance $C_{L}$ will be temporarily added to evaluate the differential resistance $R_{D}$ (see section \ref{sec:rd-id}).}
	\label{fig:fig2}
	\end{figure}

A simplified electrical model for the video part of the power detector is given in Fig. \ref{fig:fig2}.

$R_{D}$ is the differential resistance associated with the diode junction. $C_{0}$ is the output capacitance required to filter out the RF carrier and harmonics. Here we neglect parasitic resistances and capacitances associated with the bulk semiconductor or connections. They will not change the overall behavior of the diode and can be easily added  in a second step if needed. Futhermore we do not seek to deeply understand the physics of the semiconductor but rather to  qualify the diode as a low-level power detector. We tested two tunnel diode detectors from Herotek. The diode basic characteristics extracted from the manufacturer data-sheet are given in table \ref{tab:diode-data-sheet}.
\begin{table}[h!!!!!]
\centering
\caption{Tested tunnel diode specifications.}
\begin{tabular}{lr}
 \multicolumn{2}{c}{(at $25^{\circ}$C and for 10 $\mu$W power input)}\\
Model:			& HEROTEK DT8012\\
Serial Number:		& SN436409 and SN436410 \\
Frequency Range: 	& $8-12$ GHz\\
Minimum sensitivity $K_{\!D}$: & 800 V/W\\
Typical VSWR:			&2.5:1 \\
Typical output capacitance $C_{0}$: & 10 pF\\
Typical video resistance $R_{D}$ : & $125~\Omega$\\
\end{tabular}
\label{tab:diode-data-sheet}
\end{table} 

The RF voltage applied to the diode junction is rectified by the non-linear current-voltage characteristic of the device. Dropping the ac-components filtered by $C_{0}$, the rectification process is modelised by the low-frequency current generator $I_{D}$ complex function of the RF power. The intrinsic fluctuations of the diode current is accounted for by the current noise source $i_{d}$. For an ideal detector, owing to the shot effect, the average current $I_{D}$ flowing in the diode junction is affected by a fluctuation of power spectral density:
\begin{equation}
S_{i_{d}}^{shot}= 2 q I_{D}
\end{equation} 
For $P_{0}=-20$ dBm,  $I_{D}$ is of the order of 100 $\mu$A (see section \ref{sec:rd-id}). The shot noise is characterized by a noise current density
 such as $i_{d}^{shot}\approx 6$ pA/$\sqrt{\mathrm{Hz}}$.  The load resistance $R_{L}$ is associated to the thermal current noise source $i_{l}$ with a PSD given by:
\begin{equation}
S_{i_{l}}= \dfrac{4k_{B}T}{R_{L}}
\end{equation}
where $k_{B}$ is the Boltzmann constant and $T$ the temperature of the load resistance.\\

The intrinsic amplifier voltage and current fluctuations are accounted for by $e_{n}$ and $i_{n}$, according to the popular Rothe-Dahlke model \cite{rothe56}. Our custom video amplifier has 36 dB gain. It is based on a AD743 JFET operational amplifier characterized by a very low current noise: $i_{n} = 7$ fA/$\sqrt{\mathrm{Hz}}$. The voltage noise has been measured with a $50 ~\Omega$ resistance at the amplifier inuput. The white voltage noise is $e_{n}=5$ nV/$\sqrt{\mathrm{Hz}}$. All measurements presented here have been realized with a high load resistance, i.e. $R_{L}=1$ M$\Omega$: thus $R_{L}\gg R_{D}$. 
Assuming a noise free incoming RF signal and a shot noise limited detector, the detected voltage PSD will be:
\begin{equation}
S_{\delta V_{\!D}}(f)= S_{e_{n}}(f)+R_{D}^{2} \left [ S_{i_{d}}^{shot}(f)+S_{i_{l}}(f)+S_{i_{n}}(f)\right ]
\label{equ:noise-budget}
\end{equation}

For the single channel measurement set-up depicted in Fig. \ref{fig:fig1} as $R_{D} \approx 100~\Omega$, the contributions of the current noise sources are negligle. The detected noise is mainly limited by the video-amplifier input voltage noise source. For $P_{0}=-20$ dBm and $K_{D}=1,000$, the white noise floor set by the detector is:
\begin{subeqnarray}
 S_{\alpha}(f)&=& - 132 \mathrm{~dB/Hz} \\
   \cr 
  S_{\delta P}(f)&=&2.5\times10^{-23} \mathrm{~W}^{2}\mathrm{/Hz}
 \end{subeqnarray}
which is equivalent to a NEP of $5$ pW$/\sqrt{\mathrm{Hz}}$. These last figures give only a rough estimation of the diode ability to detect small power variations. Indeed they do not report the detector intrinsic noise $S_{i_{d}}(f)$ and especially its possible flicker (or $1/f$) component.

\section{Diode characteristics measurements}
\label{sec:diode-carac}
\subsection{Experimental set-up}
Two identical DT8012 diodes have been tested. These were new components that have never been used before. The microwave signal is distributed to these diodes by a Wilkinson power divider. This assembly is fasten on the $2^{\mathrm{nd}}$ stage of a Pulse-Tube cryocooler. Three low thermal conductivity coaxial semi-rigid cables connect this set to  feedthroughs placed on the cryostat top flange. The room temperature measurements have been taken the crycooler off. When on the $2^{\mathrm{nd}}$ stage temperature stabilizes at $4$ K in less than 20 h.

\begin{figure}[ht!]
	\centering
	\includegraphics[width=0.45\columnwidth]{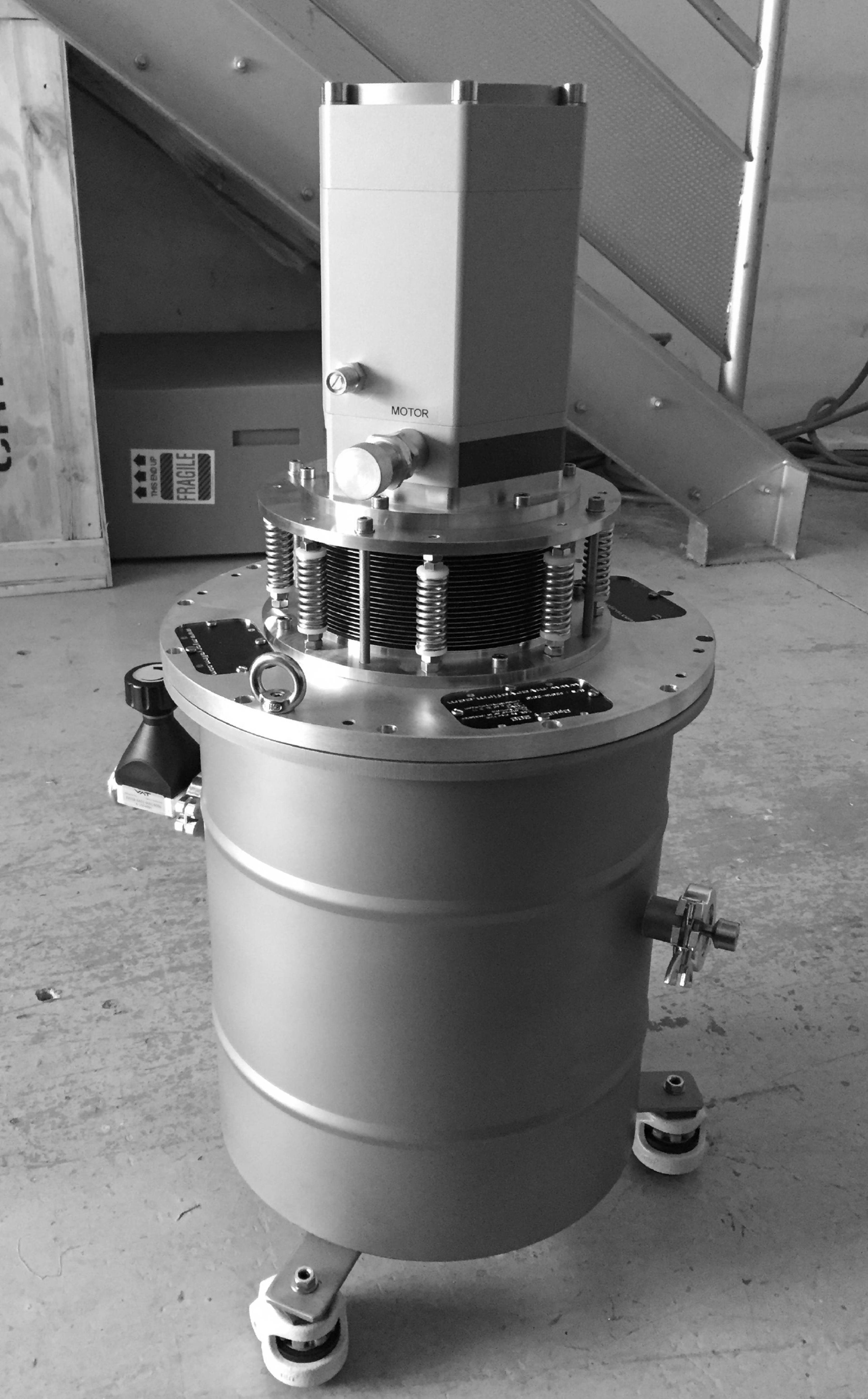}
	\caption{Cryocooled cryostat used to cool down the tunnel diodes to 4 K.}
	\label{fig:setup}
	\end{figure}

\subsection{Power-to-Voltage gain}

\begin{figure}[ht!]
	\centering
	\includegraphics[width=0.85\columnwidth]{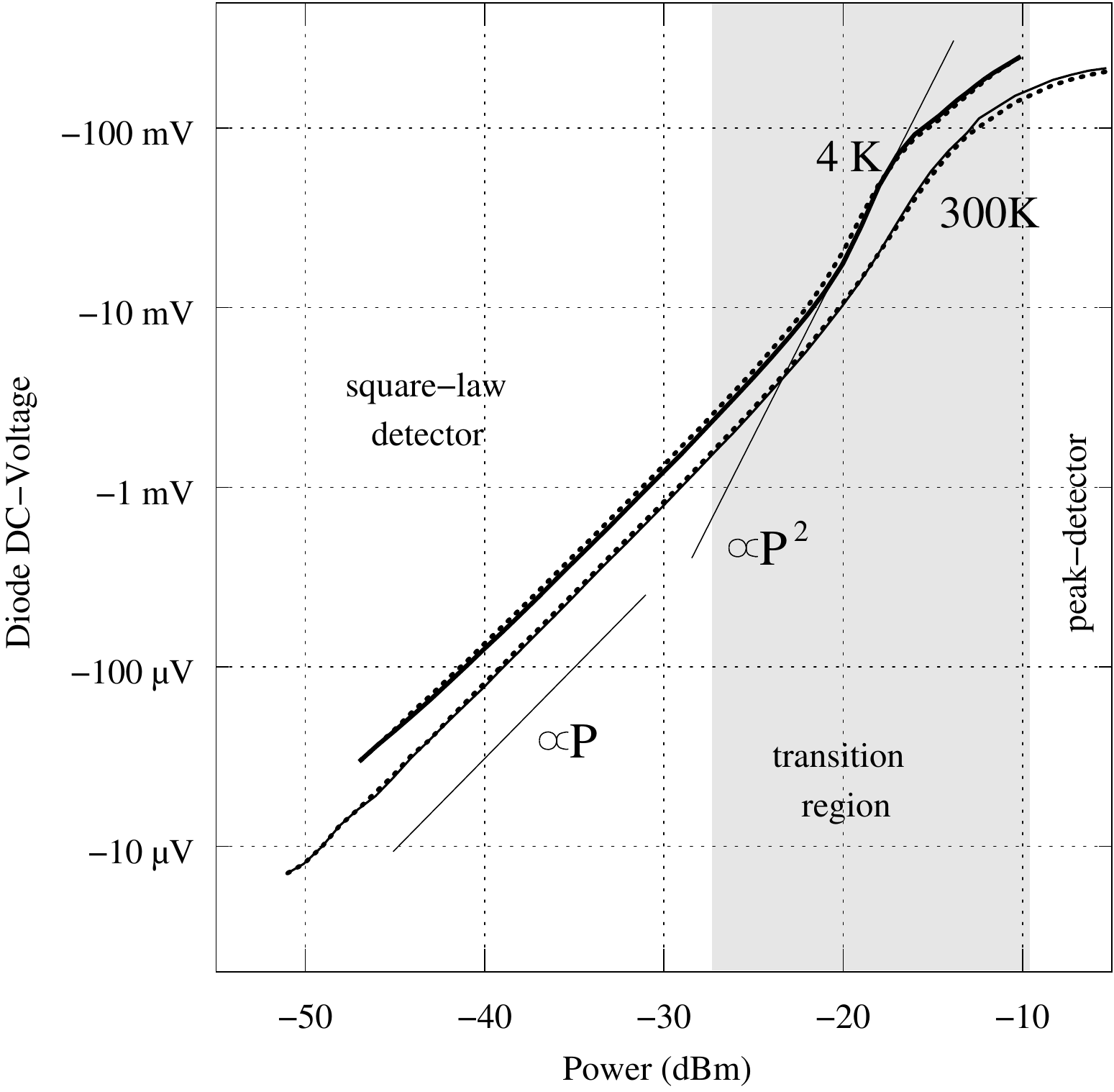}
	\caption{Power-to-Voltage gain of the tunnel diodes measured at room temperature and 4 K. The signal frequency is $\nu_{0}=10~$GHz, the amplifier input resistance is $R_{L}\!\!=\!1~$M$\Omega$.  Solid line: SN436409. Dotted line:  SN436410.}
	\label{fig:fig4}
	\end{figure}

The measurement of $K_{\!D}$ is straighforward. The video amplifier output voltage $V_{out}$ is recorded as a function of the input RF power. The result is eventually divided by the dc-gain of the video amplifier. The characteristics recorded at room temperature and 4 K are given in Fig. \ref{fig:fig4}

We can distinguish three different types of operation. At very low input power, i.e. $P_{0}\! \le \! -30$ dBm,  the diode behaves like a perfect square-law detector. Its output dc-voltage is proportionnal to the input power as expected. The sensitivity $K_{D}\approx 850$ V/W at room temperature and increases up to 1200 V/W at 4 K.  At high input power, the diode acts like a peak detector. In between, the voltage-to-power slope indicates that the output voltage depends on the square of the incoming signal power. It is obvious that equation \ref{equ:VD} is no valid in this region and its application can induce serious inaccuracy in power measurements. The same behavior has been observed in room temperature Schottky diode detectors \cite{harrison94}. In this case it has been demonstrated that this behavior is inherent to the diode exponential current-voltage characteristic. Tunnel diode characteristic does not follow the Schottky diode equation but an exponential form can be derived as well \cite{demassa70,demaagt92}. In our applications the RF signal power is often in the transition region, and thus  we need to clarify the detector sensitivity.

\subsection{Differential gain}
 Equation \ref{equ:VD} should be replaced by a more complex function of the incoming power such as $V_{D}=k(P)$. In radiometric measurements the detector transfer function should be know in its whole dynamic range.
Various methods are used to account for the nonlinear detector response \cite{hoer76,reinhardt95,walker04}. In the case of AM noise measurement or control, we are only interested in small power fluctuations $\delta P$ around the mean value $P_{0}$, with $\delta P\! \ll \! P_{0}$. Dropping constant terms and keeping only the first order term in the Taylor development, the output voltage fluctuations can be written as:
\begin{equation}
\delta V_{D}= k'(P_{0})~ \delta P + \mathcal{O}(\delta P^{2}) 
\label{equ:Taylor}
\end{equation}

\begin{figure}[h!!!!]
	\centering
	\includegraphics[width=\columnwidth]{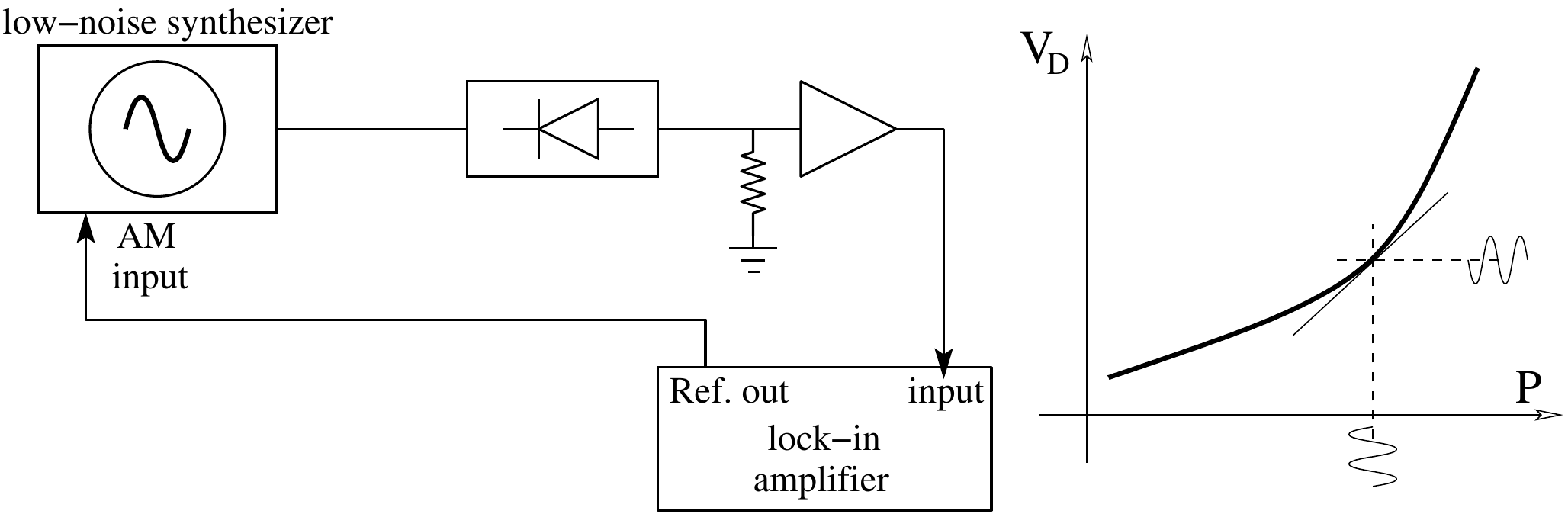}
	\caption{Differential gain measurement. The synthesizer is a Keysight E-8257D, the lock-in amplifier a  Standford Research SR810.}
	\label{fig:measurement-ac-sensitivity}
	\end{figure}

We define the first-derivative of the function $k(P)$ evaluated at $P=P_{0}$ as the differential gain of the diode. $k'(P_{0})$ can be easily measured by using the set-up depicted in Fig. \ref{fig:measurement-ac-sensitivity}.

\begin{figure}[h!!!!]
	\centering
	\includegraphics[width=\columnwidth]{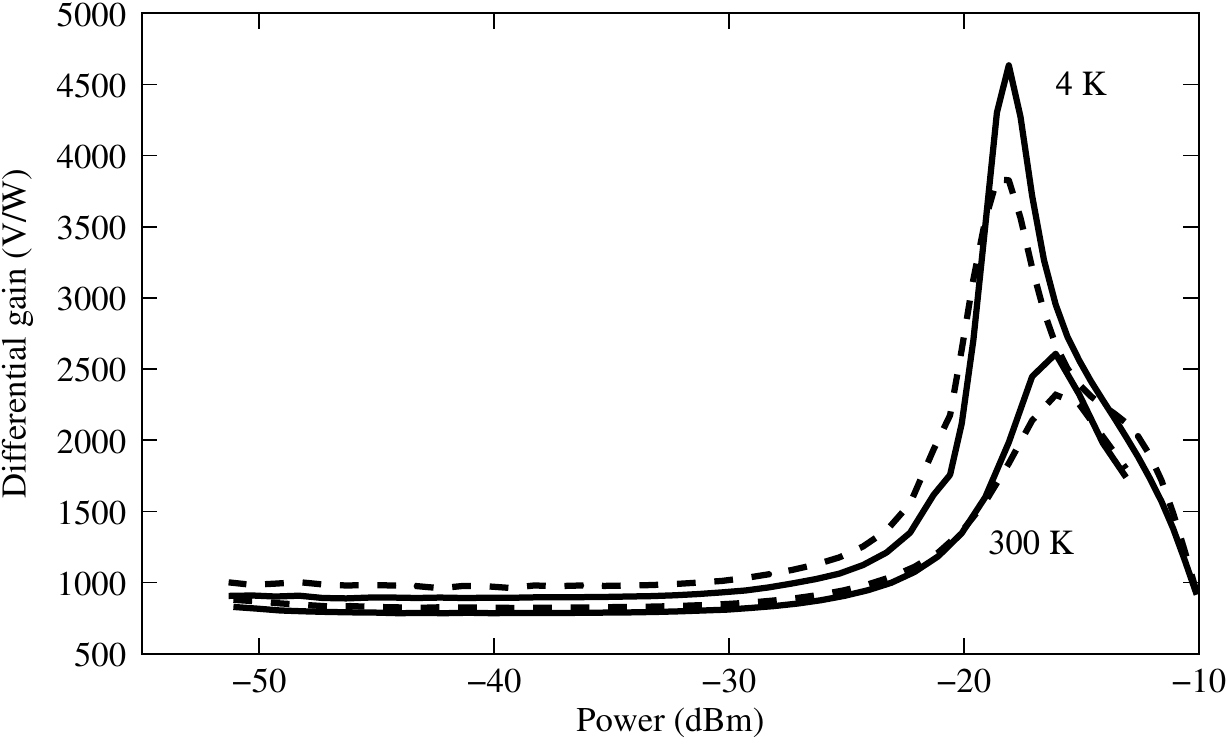}
	\caption{Tunnel diode differential gain measured at room temperature and 4 K. The signal frequency is $\nu_{0}=10~$GHz, the amplifier input resistance is $R_{L}=1~$M$\Omega$.  The modulation frequency was 70 kHz, the modulation index was $2\times10^{-2}$. Solid lines: SN436409. Dashed lines:  SN436410.}
	\label{fig:sensibilite-AC-diode-tunnel-4K-300K-dB-20150919}
	\end{figure}

We intentionnaly modulate the signal amplitude with a known small deviation and detecte with a lockin amplifier the resulting ac-voltage at the detector ouput. The measured differential gain as a function  of incoming power is represented in Fig. \ref{fig:sensibilite-AC-diode-tunnel-4K-300K-dB-20150919}.

At a very low power the differential gain is almost constant and compatible with the previous  measurements. However it shows a dramatic increase in the transition region corresponding to the change in slope in the figure \ref{fig:fig4}.  At 4 K and around  $20~\mu$W the diode is approximately four times as sensitive as at low power. Then above $\sim -13$ dBm the differential gain rapidly drops. The change in sensitivity is less abrupt at room temperature but still exists. This change should be absolutly taken into account in the following to determine with accuracy the noise floor of the AM detection.\\

\begin{figure}[ht!]
	\centering
	\includegraphics[width=\columnwidth]{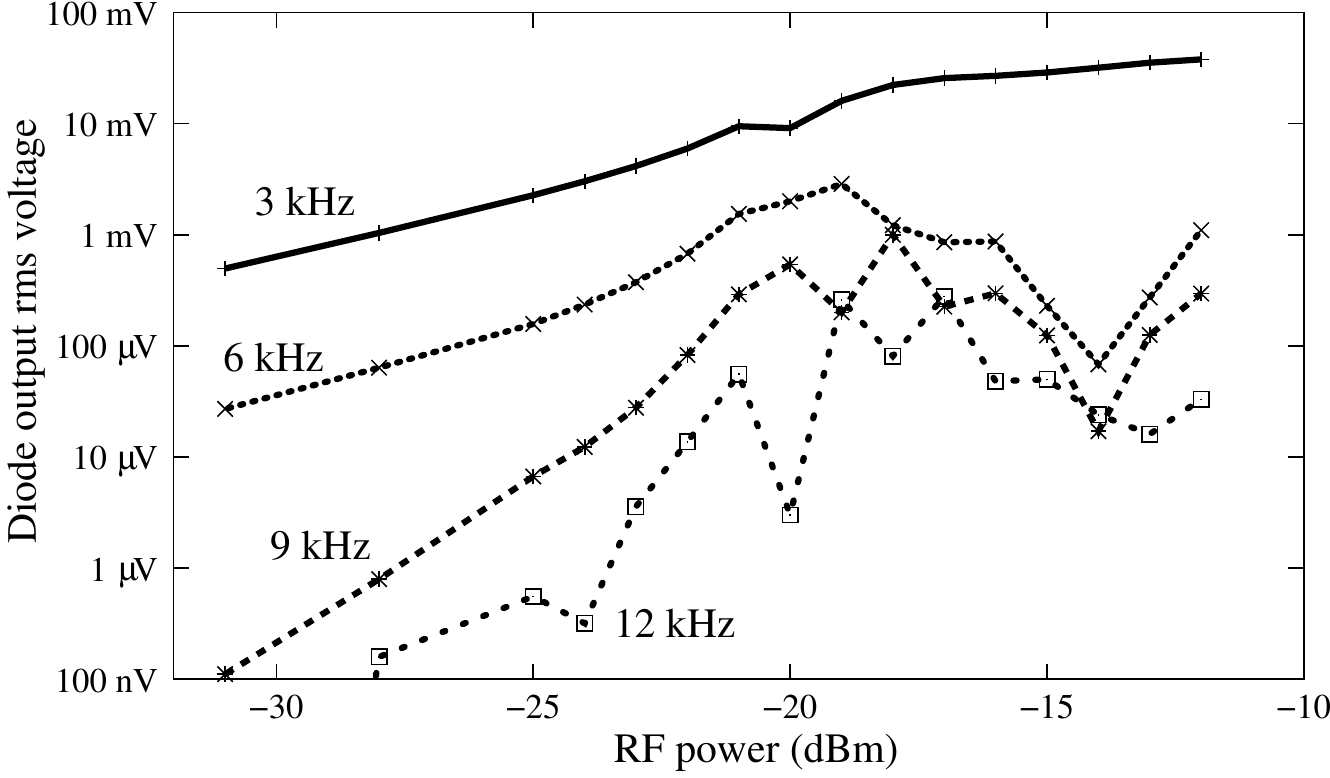}
	\caption{Harmonics in the diode output voltage.The RF signal frequency is $\nu_{0}=10~$GHz, the amplifier input resistance is $R_{L}=1~$M$\Omega$. The AM frequency was 3 kHz, the modulation index was $2\times10^{-1}$. Diode SN436409.}.
	\label{fig:harmonic}
	\end{figure}

The non-linear behavior of the detector is also attested by the presence of harmonics of the AM frequency at the diode output. These harmonics come from the higher order terms in the development of $k(P)$ neglected in equation \ref{equ:Taylor}. The figure \ref{fig:harmonic} gives the level of harmonics for a AM frequency of 3 kHz and a modulation index of 0.2.

\subsection{Differential resistance and diode current}
\label{sec:rd-id}
\begin{figure}[h!!!!]
	\centering
	\includegraphics[width=\columnwidth]{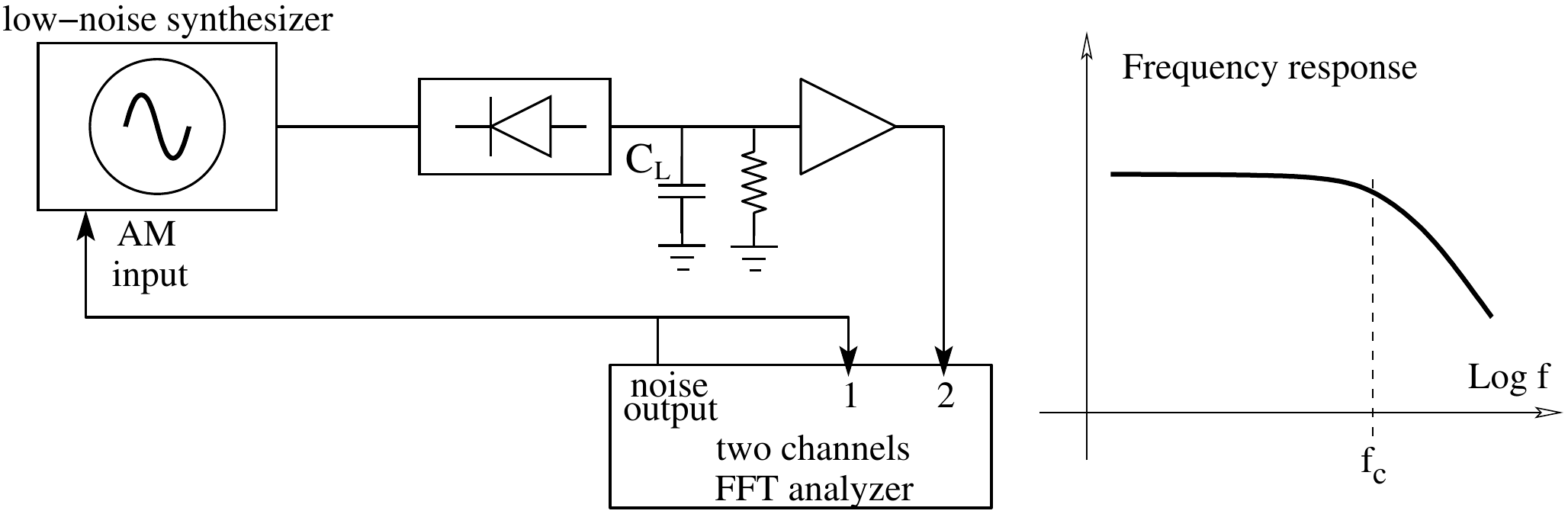}
	\caption{Video bandwidth measurement. The added capacitance is $C_{L}=100$ nF.The synthesizer is a Keysight E-8257D.}
	\label{fig:measurement-fc}
	\end{figure}

The diode differential resistance $R_{D}$ appears in the detector noise budget given by the equation \ref{equ:noise-budget}. We determine its variation as a function of the RF signal power with the set-up schematized in the figure \ref{fig:measurement-fc}.
	
\begin{figure}[ht!]
	\centering
	\includegraphics[width=\columnwidth]{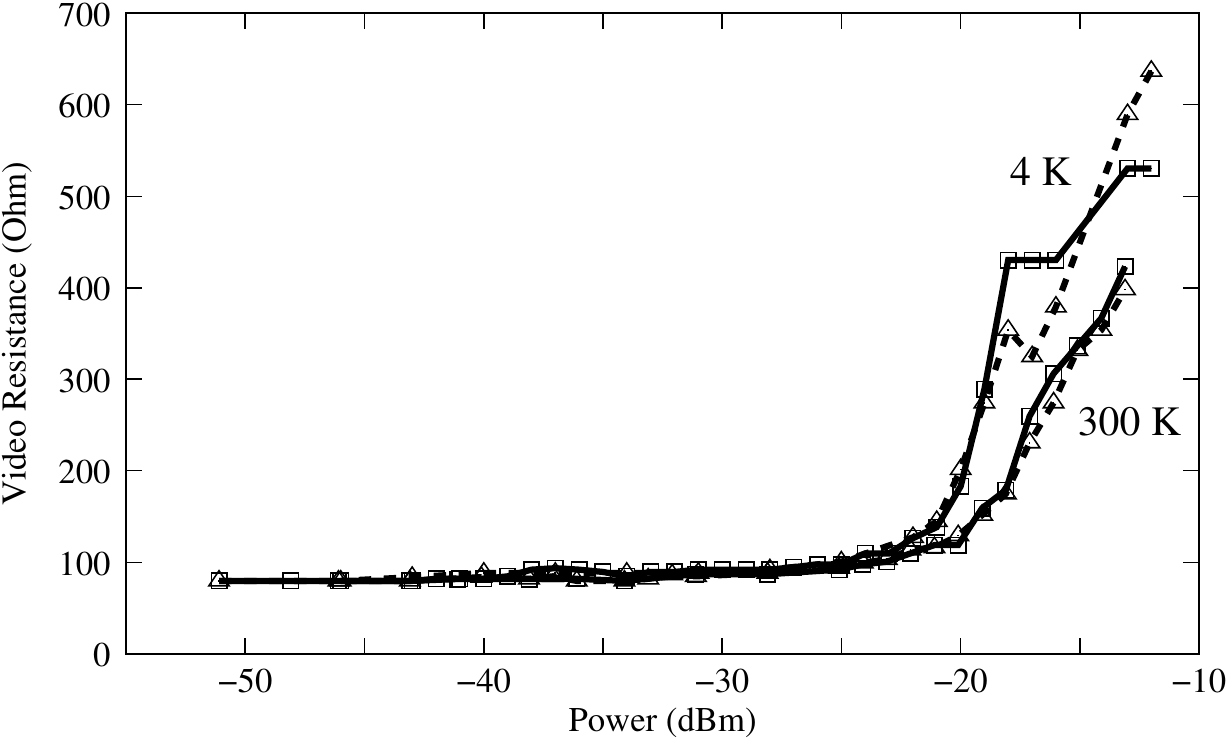}
	\caption{Diode differential resistance $R_{D}$ as determined from the measurement of the video bandwith in presence of the additional capacitor $C_{L}=100$ nF. Solid lines: SN436409. Dashed lines:  SN436410.}
	\label{fig:video-resistance-4K-300K-dB}
	\end{figure}

In the simplest configuration depicted in the figure \ref{fig:fig1}, the video bandwidth is larger than 100 kHz. Indeed the high load resistance ($R_{L}=1$ M$\Omega$) is shorted by the diode differential resistance: $R_{L} \gg R_{D}$.  By adding a large capacitance $C_{L}=100$ nF, the cutoff frequency $f_{C}\approx (2 \pi R_{D}C_{L})^{-1}$ is shifted below 100 kHz. $f_{C}$ can be easily measured with a baseband Fast Fourier Transform analyzer. Fig. \ref{fig:video-resistance-4K-300K-dB} shows $R_{D}$ as a function of the signal power.\\

\begin{figure}[ht!]
	\centering
	\includegraphics[width=\columnwidth]{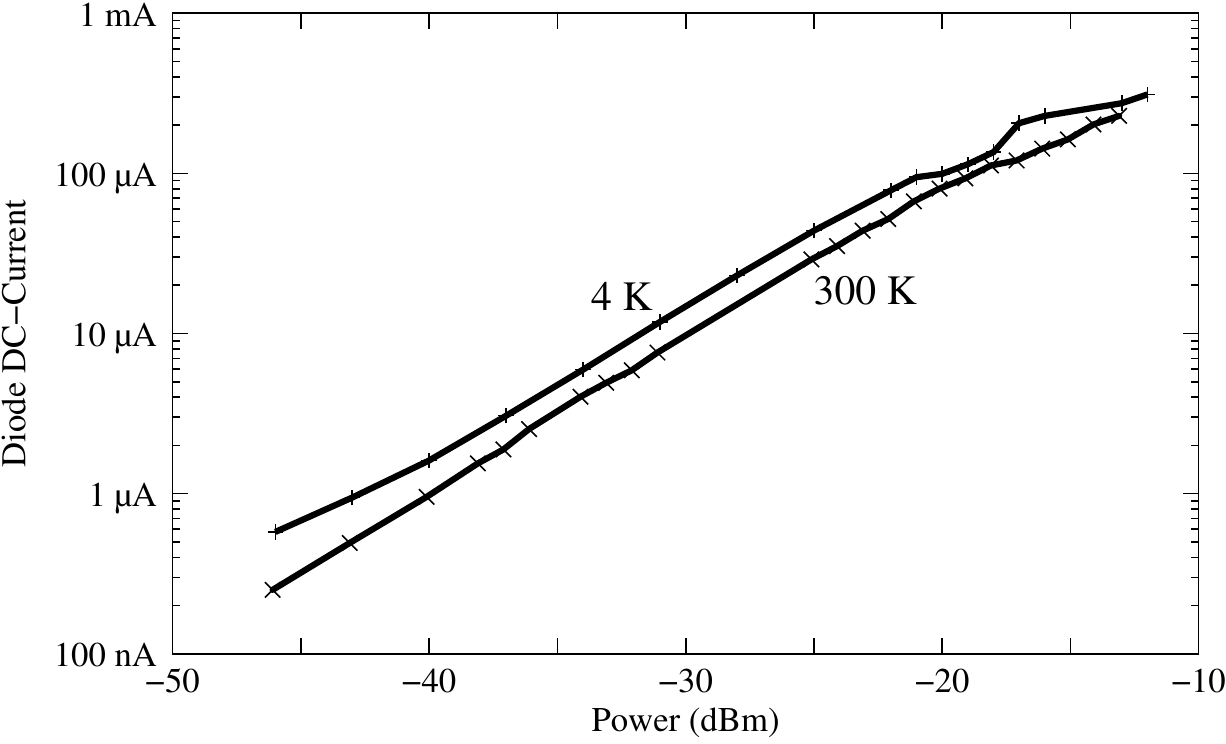}
	\caption{Diode current as a function of the signal power. Diode SN 436410.}
	\label{fig:current}
	\end{figure}

We note the transition region is characterized by an strong increase of $R_{D}$.
The diode current $I_{D}$ can now be calculated. As $R_{L}\gg R_{D}$, the output voltage is: $V_{D}=R_{D}\times I_{D}$.
The diode current is given in Fig. \ref{fig:current}.

 At low input power $I_{D}$ is proportional to the incoming power. The \it{responsivity}\rm\ is $8.4$ A/W at 300K and $10.8$ A/W at 4K. The diode current variation is smoother than those of the other parameters. Fig. \ref{fig:current} suggests that extended linearity and dynamic range can be obtained by loading the output to a virtual ground or to other low-impedance circuit. 

\section{Noise measurement}
\label{sec:diode-noise}	
\begin{figure}[h!!!!!]
	\centering
	\includegraphics[width=\columnwidth]{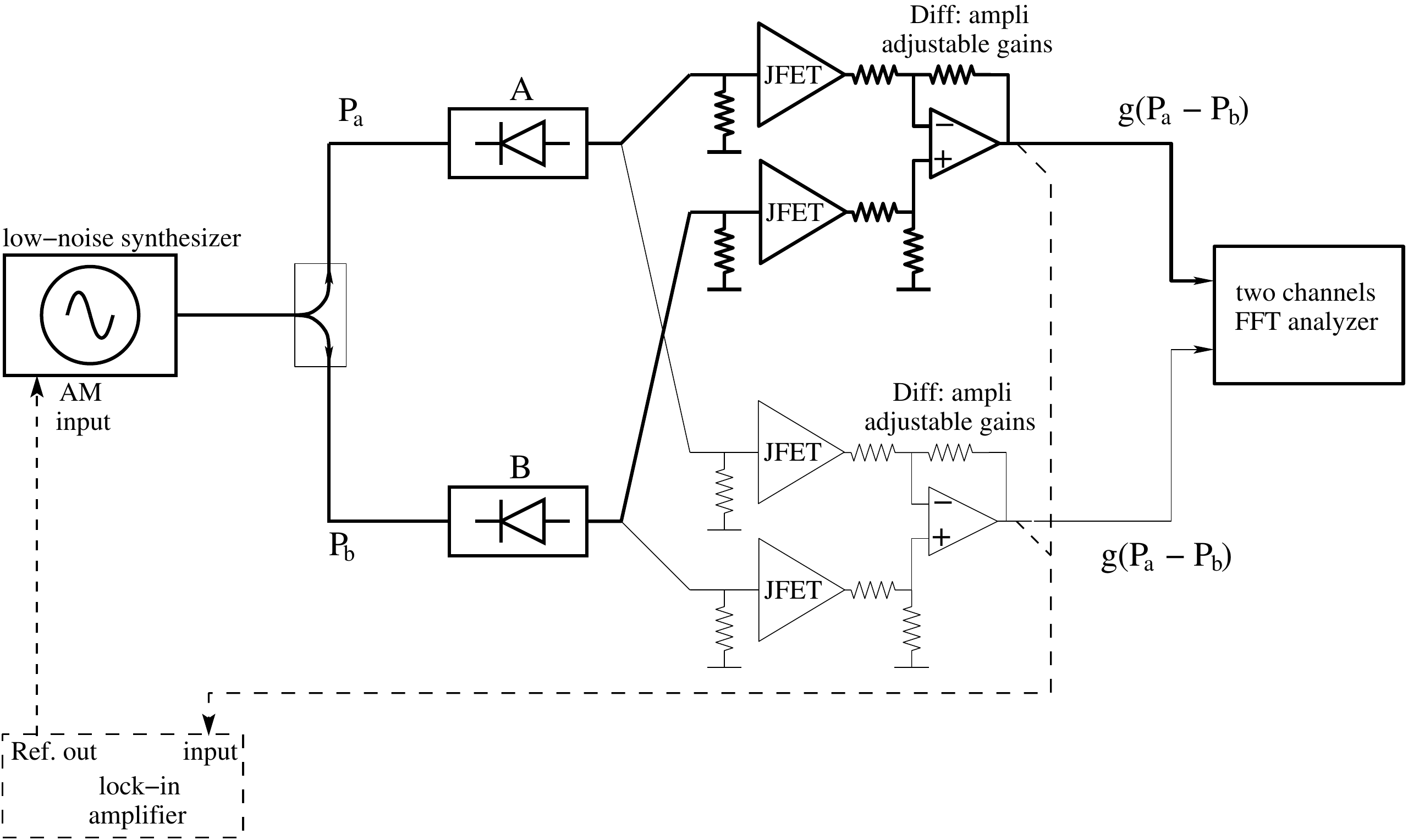}
	\caption{Two-channels correlation measurement set-up. The lockin amplifier is used to adjust the gains of the differential amplifiers. 
	}
	\label{fig:two-detector-correlation-measurement}
	\end{figure}

A detector alone can be measured only if a reference source is available whose AM noise is lower than the detector noise, and if the amplifier noise can be made negligible. These are unrealistic requirements. 
 A more sophistical approach is to compare two detectors, A and B as shown in the figure \ref{fig:two-detector-correlation-measurement}.\\

\begin{figure}[ht!]
	\centering
	\includegraphics[width=\columnwidth]{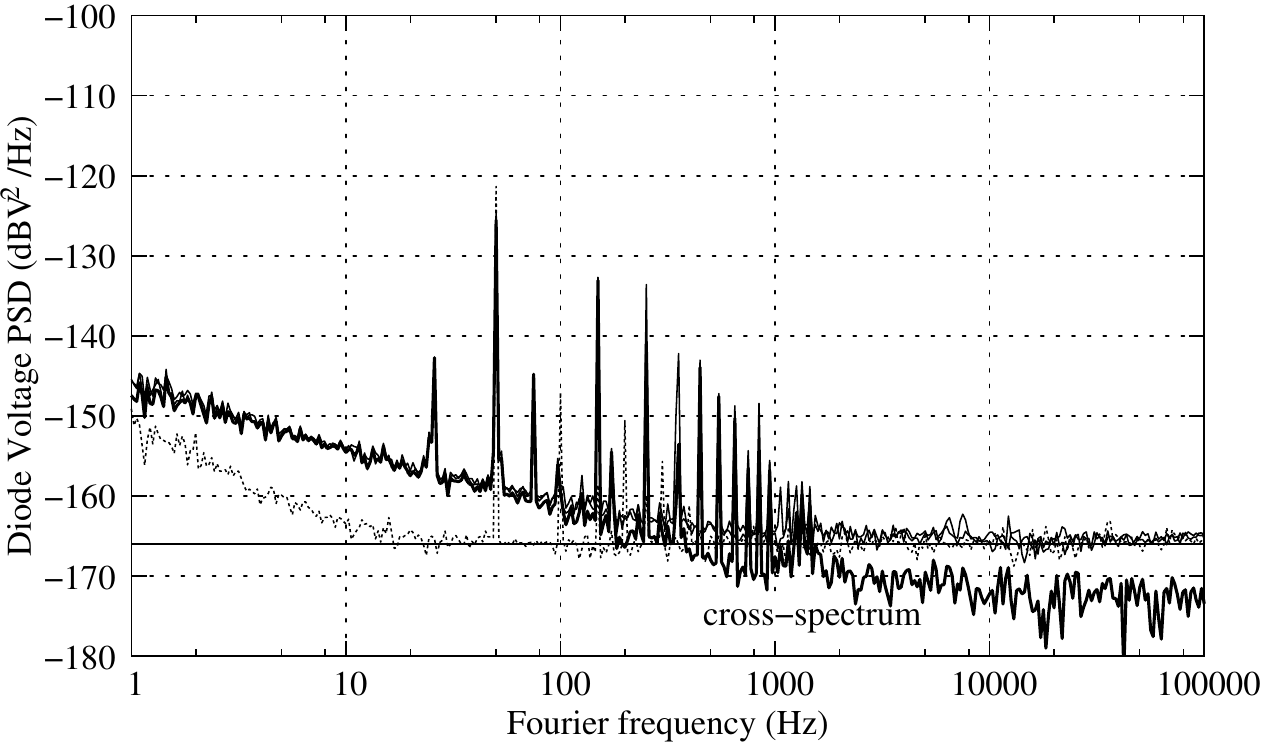}
	\caption{Diode voltage PSD at 4 K and -20 dBm signal power. Thin lines: PSD of each chanel. Bold line: Cross-spectrum. Dashed line: Single chanel noise floor.}
	\label{fig:Voltage-PSD-Psynthe-11.1dBm}
	\end{figure}

The first arm (bold line) takes the differential signal $g(P_{b} - P_{a}) \approx 0$, which is not affected by the power fluctuation of the source. The voltage PSD measured at the output still contains the contribution of the video amplifiers noise sources. Thus a second arm identical to the first one measures also $g(P_{b} - P_{a})$. By measuring the cross spectrum at the output of two arms we eliminate the contribution of the intrinsic voltage noise source $e_{n}$ of each video amplifier. The noise current $i_{n}$ of the video-amplifier turns into correlated random voltage fluctuations across each diode video resistance, and thus is not eliminated by correlation. Using a JFET operational amplifier makes the current noise contribution negligible. The figure \ref{fig:Voltage-PSD-Psynthe-11.1dBm} shows the typicall diode voltage PSD obtained with the two diodes at 4 K and receiving $-20$ dBm. We assume the two detectors equivalent: 3 dB have been substracted from the measured spectrum.

The video-amplifier voltage noise $e_{n}$ has been measured with a $50~\Omega$ input load (dashed line): $e_{n}=5$~nV/$\sqrt{\mathrm{Hz}}$. On each channel this noise floor masks the diode intrinsic noise for Fourier frequencies higher than  $1$ kHz. The later is revealed by the cross-spectrum (bold line). The diode noise is dominated by a flicker ($1/f$ slope) component and a white noise floor reaching $\sim -173$ dBV$^{2}/$Hz for $f>30$ kHz.\\

It is useful to express the intrinsic diode noise in term of $S_{\delta P}(f)$ (Power fluctuations PSD) or $S_{\alpha}(f)$ (relative amplitude fluctuations PSD). Combining equations \ref{equ:Salpha} and \ref{equ:Taylor}, they are related to the experimenal measurement of diode voltage PSD as:
\begin{subeqnarray}
  S_{\delta P}(f) &=&{\left [{ \dfrac{1}{k'(P_{0})}} \right ]}^{2} S_{\delta V_{d}}(f)   \\
  \cr 
S_{\alpha }(f) &=& {\left [{ \dfrac{1}{2 P_{0}k'(P_{0})}} \right ]}^{2} S_{\delta V_{d}}(f)
 \end{subeqnarray}

\begin{figure}[ht!]
	\centering
	\includegraphics[width=\columnwidth]{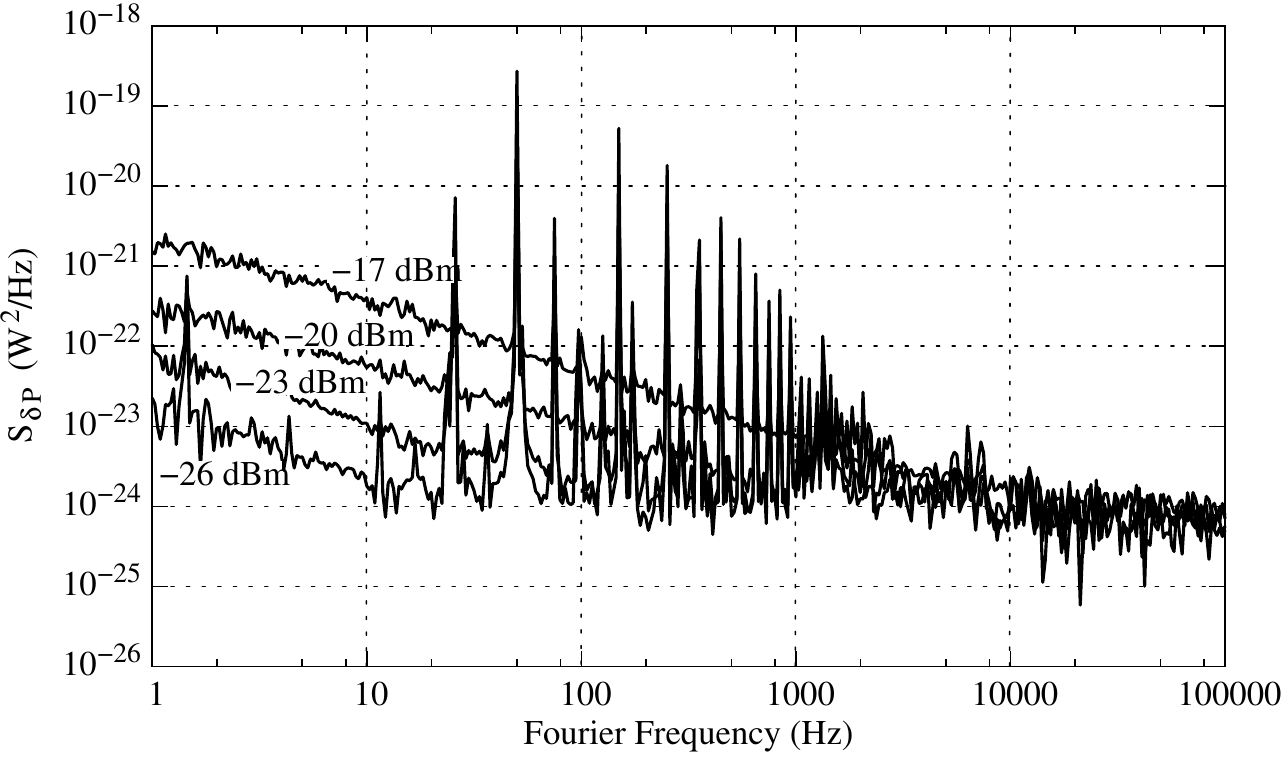}
	\caption{Diode intrinsic noise expressed as the input power fluctuations PSD $S_{\delta P}(f)$ at 4 K for  -26,-23, -20 and -17 dBm input signal power.}
	\label{fig:power-PSD-some}
	\end{figure}

\begin{figure}[ht!]
	\centering
	\includegraphics[width=\columnwidth]{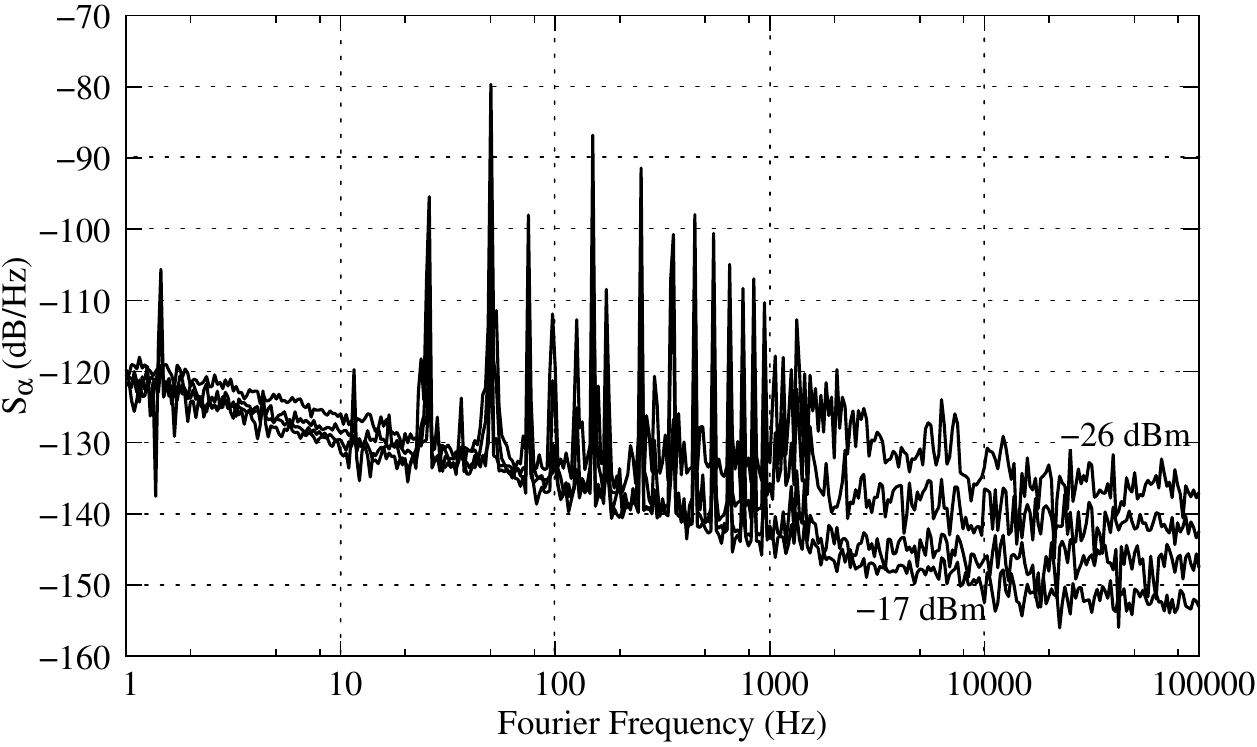}
	\caption{Diode intrinsic noise expressed as the relative amplitude fluctuations PSD of the incoming RF signal. $S_\alpha(f)$  is measured at 4 K for  -26,-23, -20 and -17 dBm input signal power.}
	\label{fig:Salpha-some}
	\end{figure}

$S_{\delta P}(f)$ and  $S_{\alpha}(f)$ are presented in Fig. \ref{fig:power-PSD-some} and Fig. \ref{fig:Salpha-some} respectively.

To keep the figures readable, we only present the PSDs for four signal powers. The same measurements have been done at room temperature. All the recorded crosspectrum show a white noise floor and flicker component, we approximate the two PSDs as:
\begin{equation}
  S_{\delta P}(f) = \dfrac{p_{-1}}{f} + p_{0}~~\mathrm{and}~~S_{\alpha }(f) =  \dfrac{a_{-1}}{f} + a_{0}   
 \end{equation}

\begin{figure}[ht!]
	\centering
	\includegraphics[width=\columnwidth]{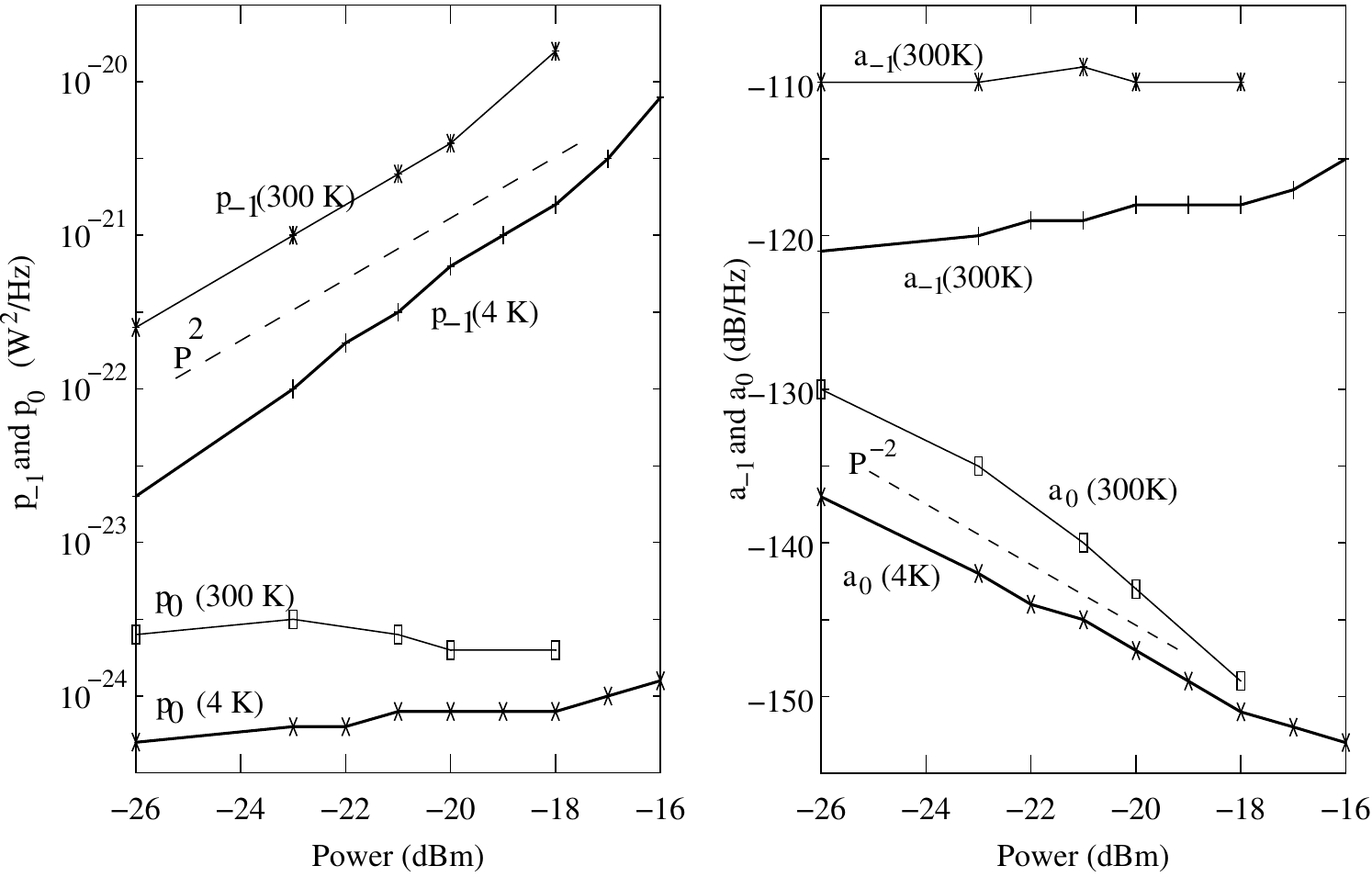}
	\caption{White and flicker noise components for $S_{\delta P}$ and $S_{\alpha}$ measured at 4 K and 300 K.}
	\label{fig:SDP-flicker-white-noise-4K-300K}
	\end{figure}

The figure \ref{fig:SDP-flicker-white-noise-4K-300K} gives the coefficients $p_{-1},~p_{0},~a_{-1}$ and $a_{0}$ as function of the signal power at 4 K and 300 K.\\
	
In term of power fluctuations PSD, the white noise floor is independant of the signal power. For $f>10$ kHz, $S_{\delta P}=6.3\times 10^{-25}$ W$^{2}/$Hz, equivalent to a NEP of $0.8$ pW$/\sqrt{\mathrm{Hz}}$. The ficker noise increases proportionally to $P^{2}$. The same dependancy has been reported in \cite{eng61}. When expressed as $S_{\alpha}$, the flicker noise appears independant of the incoming power. The relative amplitude fluctuations PSD is about $-120$ dB/Hz at 1 Hz. The room temperature measurements are about $10$ dB higher.

\section{Application to power control}
\label{sec:appli}
\begin{figure}[ht!]
	\centering
	\includegraphics[width=\columnwidth]{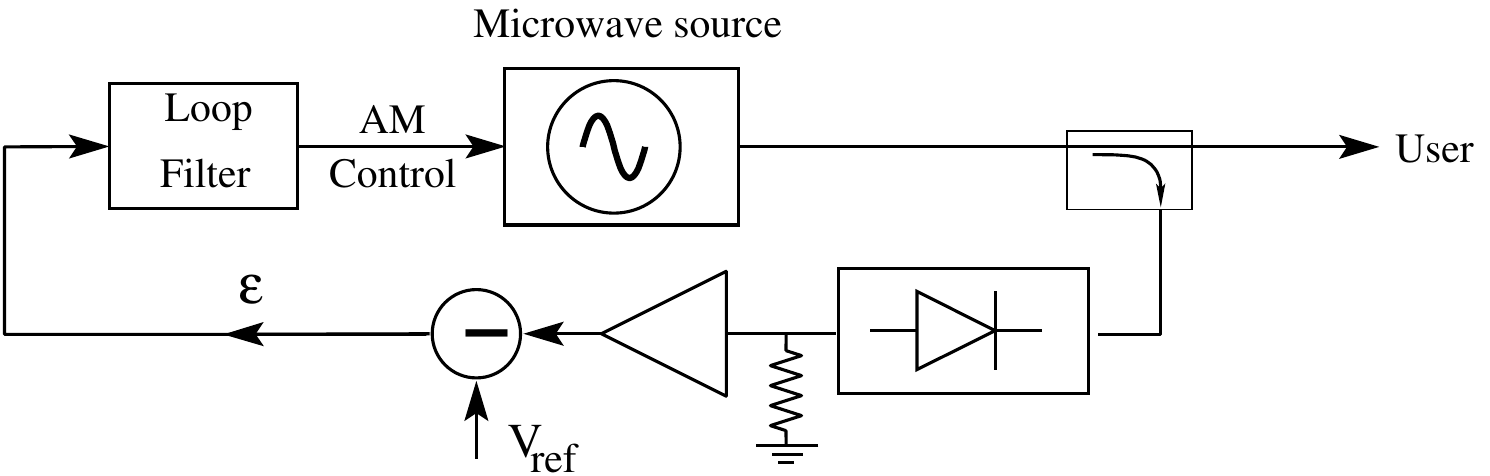}
	\caption{Power control principle.}
	\label{fig:P-control}
	\end{figure}

The last results are useful to evaluate the ultimate limit of a power control using such a tunnel diode as detector. Let us assume a classical power control set-up based on the scheme given in the figure \ref{fig:P-control}.

The amplified diode voltage is compared to a stable voltage reference. The obtained error signal $\epsilon$ is integrated by the loop filter to be applied to the electronic amplitude control of the source. Assuming a loop bandwidth of 100 Hz or so, the residual amplitude fluctuations will ultimatly given by the diode flicker noise independant of the incoming microwave power:
\begin{equation}
S_{\alpha}(f)=\dfrac{a_{-1}}{f}
\end{equation}
In this case the fractional amplitude stability is independant of the integration time $\tau$. The Allan standard deviation is \cite{rubiola-phase-noise}:
\begin{equation}
\sigma_{\alpha}(\tau)=\sqrt{2\ln(2)a_{-1}}
\end{equation}

For a cryogenic diode the fractional amplitude stability limit set by the diode flicker noise is $1.2\times 10^{-6}$. For a room temperature diode the limit is $3.7\times10^{-6}$. These ultimate limits will obviously be degraded at long integration time by the thermal sensitivity of voltage reference and of the diode itself. 
\section{Acknowledgments}

The work has been realized in the frame of the ANR project: Equipex Oscillator-Imp. The authors would like to thank the Council of the R\'egion de Franche-Comt\'e for its support to the Projets d’Investissements d'Avenir.

\end{document}